\documentstyle[graphicx,aps]{revtex}
\begin{document}
\title{Synthesis and characterization of the infinite-layer superconductor 
Sr$_{0.9} $La$_{0.1}$CuO$_{2}$ }
\author{C. U. Jung\footnote{Electronic address: jungking@postech.ac.kr}, J. Y. Kim, Mun-Seog Kim, S. Y. Lee, and Sung-Ik Lee}
\address{National Creative Research Initiative Center for Superconductivity 
and Department of Physics, Pohang University of Science and Technology,
Pohang 790-784, Republic of Korea}
\author{D. H. Ha}
\address{Korea Research Institute of Standards and Science,
PO Box 102, Taejon 305-600, Republic of Korea}
\maketitle

\begin{abstract}
We report the high-pressure synthesis of the electron-doped infinite-layer
superconductor Sr$_{0.9}$La$_{0.1}$CuO$_{2}$. A Rietveld analysis using
X-ray powder diffraction data showed that, within the resolution of the
measurement, the sample was purely an infinite-layer structure without any
discernible impurities. The superconducting volume fraction and the
transition width were greatly improved compared to those in the previous
reports. Also the irreversibility field line was much higher than that of
(La,Sr)$_{2}$CuO$_{4}$. The higher value seems to originate from the strong
interlayer coupling due to the reduced average distance between the CuO$_{2}$
planes.
\end{abstract}

\section{Introduction}
The electron-doped infinite-layer compounds (Sr$_{1-x}^{+2}$Ln$_{x}^{+3}$)CuO%
$_{2}$ (Ln = La, Sm, Nd, Gd, etc.) consist of an infinite stacking of CuO$%
_{2}$ planes and metallic (Sr) layers.\cite{Siegrist88} This is the simplest
structure that contains only the key ingredients of all high-$T_{c}$
cuprates. The charge reservoir block common to other cuprate superconductors
does not exist in this compound. Superconductivity in the infinite-layer
compounds was first observed by Smith {\it et al}. for (Sr$_{1-x}$Nd$_{x}$%
)CuO$_{2}$.\cite{Smith91} Since the structure is so simple, it provides a
unique opportunity to explore superconductivity unscreened by\ the charge
reservoir block.

Several interesting properties have been observed for infinite-layer
compounds. Since the average distance between CuO$_{2}$ planes is reduced
due to the absence of the charge reservoir block, the interlayer coupling is
expected to be very strong, which should strengthen the superconductivity.
However, $T_{c}$ is only about 43 K,\cite
{Ikeda93,Jorgensen93,Er91,Er92,Er94,Review,Yao94,Er97} and neither the size
of ionic radius, the magnetic moment, nor the concentration of Ln ions at Sr
sites affects $T_{c}$.\cite{Ikeda93} Moreover, the oxygen is very
stoichiometric; neither vacancies nor interstitial oxygen exists.\cite
{Jorgensen93}

So far, the study of superconductivity in these compounds has been hindered
by the lack of high quality bulk samples. However, in the case of
infinite-layer superconducting films, superconducting transition temperature
is much lower than those of bulk samples. For an infinite-layer
superconducting film of (Sr$_{1-x}$Nd$_{x}$)CuO$_{2}$, the $T_{c}$ is
reduced by about one half.\cite{Sugii94,Jones94} High-pressure synthesis is
a unique method that now allows infinite-layer compounds to be made in bulk
form with various lanthanide ions doped into Sr sites with larger
superconducting volume fractions.\cite
{Ikeda93,Jorgensen93,Er91,Er92,Er94,Review,Yao94,Er97}

In this paper, we report the high-pressure synthesis of\ the infinite-layer
superconductor Sr$_{0.9}$La$_{0.1}$CuO$_{2}$ (Sr(La)-112). To identify the
superconductivity, we measured the low-field magnetization. The resulting
superconducting volume fraction was found to be greatly improved compared to
the results reported so far. The high purity of the samples was confirmed by
a Rietveld analysis of the X-ray powder diffraction data. The high-field
magnetization curves, along with the resulting irreversibility line showed,
that the pinning was unusually high compared to that for hole doped cuprates.

\section{experimentals}

A cubic multi-anvil-type press was used to synthesize Sr(La)-112.\cite
{Ikeda93} The precursors were prepared by using the solid state reaction
method.\cite{Ikeda93,Er91} Starting materials of La$_{2}$O$_{3}$, SrCO$_{3}$%
, and CuO were mixed to the nominal composition of Sr$_{0.9}$La$_{0.1}$CuO$%
_{2}$. The mixture was then calcined at 950 $^{\circ }$C\ for 36 hours with
several intermittent grindings. The pelletized precursors sandwiched by Ti
oxygen getters were put into a Au capsule in a high-pressure cell. A {\it D}%
-type thermocouple was used to monitor the temperature.

The pressure cell was compressed up to 4 GPa and then heat-treated with a
graphite-sleeve heater. The temperature of the Au capsule was calibrated to
the heating power, and that data allowed us to use the heating power to
control the sample-cell temperature. However much of the power from the
power-supply was lost to the stray resistance ({\it R}$_{{\rm stray}}\sim
10^{-2}\Omega $ ) between the power-supply and the graphite heater ({\it R}$%
_{{\rm heater}}\sim 10^{-2}\Omega $ ). Even though the power was constantly
supplied from the power-supply, the actual heating power of the sample
fluctuated because {\it R}$_{{\rm heater}}$ changed during the synthesis; $%
\Delta ${\it R}$_{{\rm heater}}/${\it R}$_{{\rm heater}}\sim 0.1$. The
amount of fluctuation was roughly proportional to {\it R}$_{{\rm stray}}$/%
{\it R}$_{{\rm heater}}$. To solve this problem, we controlled the heating
power across the sample instead of the main power. With this method, a
temperature stability of $\pm $ 2 $^{\circ }$C was obtained for a two-hour
heating time under high-pressure conditions.

The heating power was increased linearly and then maintained constant for 2
hours. The synthesizing temperature was about $1000$ $^{\circ }$C. Then, the
sample was quenched to room temperature after an additional postannealing at 
$500\sim 600$ $^{\circ }$C for 4 hours. Two conditions were important in
obtaining higher quality samples. One was the long-term stability of the
synthesizing temperature, and the other was the uniformity of the
temperature inside the sample cell, the former being more important. The
pressure cell and the heating conditions were optimized after hundreds of
trials, very homogeneous samples larger than $200$ mg were obtained.

The samples were characterized by powder X-ray diffraction (XRD)
measurements using Cu {\it K}$\alpha $ radiation and a SQUID magnetometer
(MPMS{\it XL}, Quantum Design). Scanning electron microscopy (SEM) and
optical microscopy were also used. Zero-field-cooled (zfc) and field-cooled
(fc) magnetization $M(T)$ curves were measured. The structural
characterization at room temperature was carried out by using the Rietveld
refinement method to analyze the X-ray powder diffraction data. The SEM
image showed closely \ packed grains of uniform size.

\section{data and discussion}
\subsection{Structure}
It is known that the phase purity obtained from X-ray or neutron powder
diffraction does not represent the purity of a superconductor. Previously,
it was claimed, based on the XRD data, that infinite-layer compounds were in
a pure phase, but with a widely distributed superconducting volume fraction. 
\cite{Ikeda93,Jorgensen93,Er91,Er92,Er94,Review,Yao94}

Our powder XRD pattern showed an infinite-layer compound with the tetragonal
space group {\it P}4/mmm and lattice parameters $a=b=3.950$ \AA\ and $%
c=3.410 $ \AA . The value of $2\theta $ was varied from 20$^{\circ }$ to 140$%
^{\circ }$ in steps of 0.02$^{\circ }$, and the integration time was 15
seconds at each point. The Rietveld refinement program RIETAN-94 by F. Izumi
with 50 parameters was used for the analysis.\cite{Izumi}

The Rietveld refinement profile is shown in Fig. \ref{XRD}. In that
analysis, the thermal factors were assumed to be isotropic, and the
coordination of each atom was fixed. We constrained the Sr:La ratio to the
nominal stoichiometry of the precursor.\cite{Jorgensen93} The obtained
values of the lattice constants agree quite well, within 0.001 \AA , with
those obtained by neutron powder diffraction.\cite{Jorgensen93} This
refinement could not identify, within the resolution, any discernible amount
of impurities. The agreement factors, $R$, between the measured and the
calculated diffraction intensities were quite small, and goodness of fit was
excellent $S=4.0008$. The refined structural parameters are summarized in
Table \ref{refinement}.

The structural analysis of an infinite-layer compound can also give valuable
information about the doping concentration because the lattice constant is
sensitive to the doping concentration.\cite{Ikeda93} Note the opposite
behavior of the lattice constants with doping; the {\it a-}axis expands
while the {\it c}-axis shrinks. The Rietveld refinement showed that the
doping concentration in our Sr$_{1-x}$La$_{x}$CuO$_{2}$ was approximately $%
x=0.1$, which is same as the nominal composition.

\subsection{Superconducting properties}
\begin{figure}[tbp]
\begin{center}
\includegraphics[width=9cm,height=12cm]{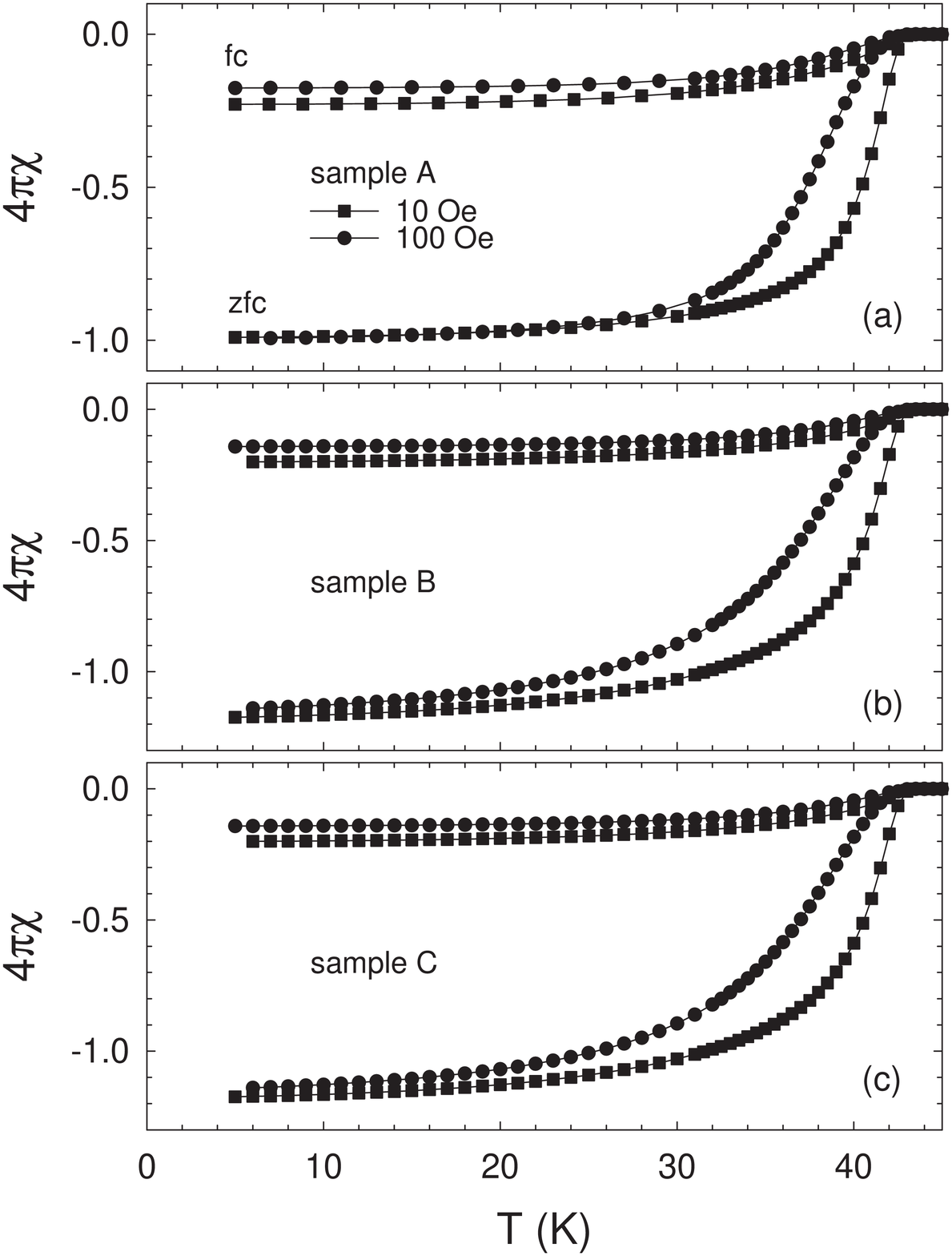}
\end{center}
\caption{Magnetic susceptibility, $4\protect\pi \protect\chi (T)$, of Sr$%
_{0.9}$La$_{0.1}$CuO$_{2}$\ for zero-field-cooling and field-cooling from
the low-field magnetization $M(T)$ at 10 and 100 Oe. For calculating the
nominal superconducting volume fraction $f_{{\rm nom}}$, we used a
low-temperature density of $5.94$ g/cm$^{3}$ from Ref. 10. (a) Sample A, $%
f=100$\%, (b) Sample B, $f=117$\%, and (c) Sample C, $f=122$\%. }
\label{lowfieldMT}
\end{figure}

\begin{figure}[tbp]
\begin{center}
\includegraphics[width=9cm,height=7cm]{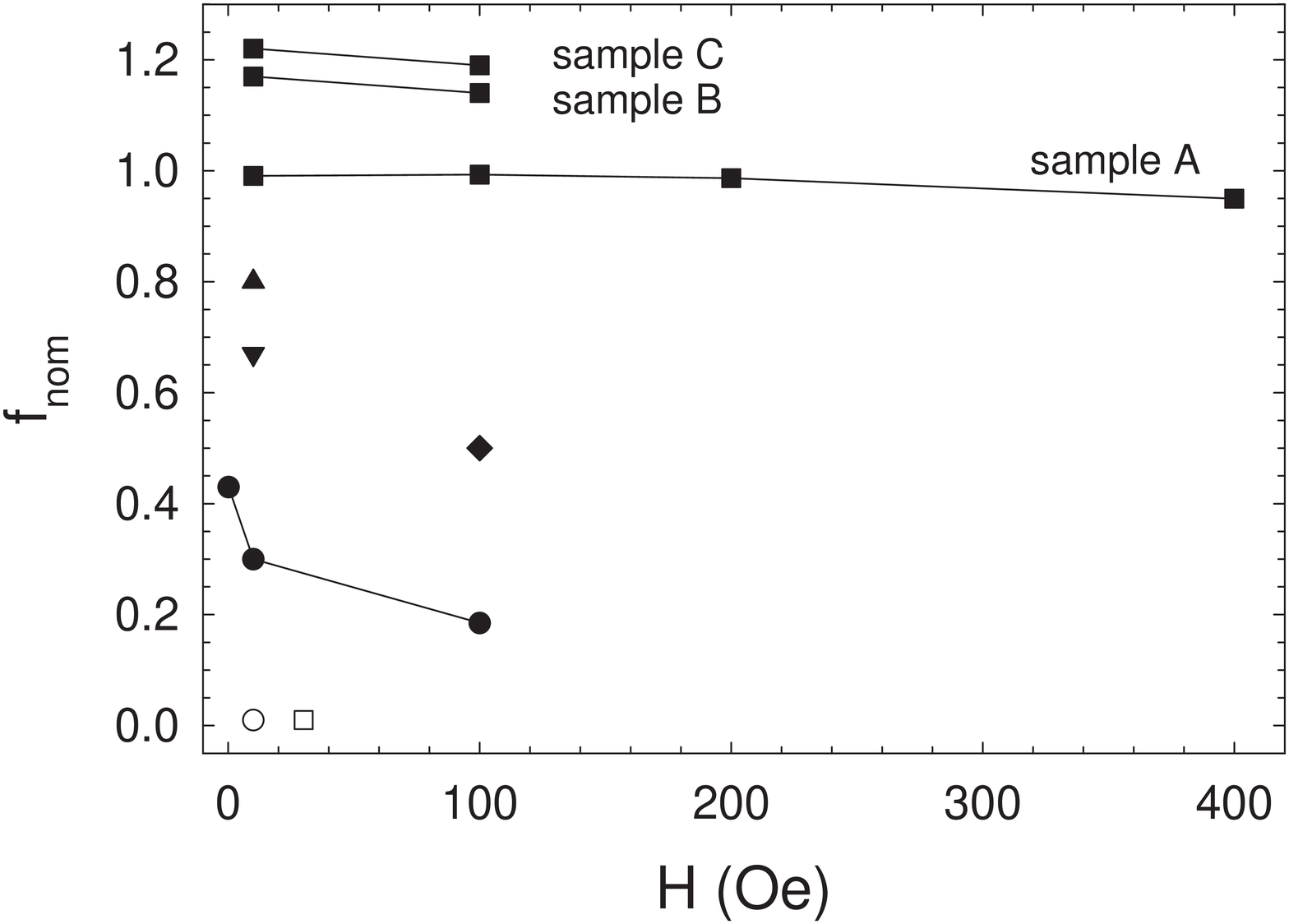}
\end{center}
\caption{Comparison of the nominal superconducting volume fraction {\it f}$_{%
{\rm nom}}=-4\protect\pi \protect\chi (T\ll T_{c})$. The squares labeled
with sample A, sample B, and sample C are data for our samples. The
up-triangle is from Ref. 7, the down-triangle is from Ref. 4, the circles
are from Ref. 6, the diamond is from Ref. 10, the open circle is from Ref.
14, and the open square is from Ref. 15. }
\label{fnomcomparison}
\end{figure}

Typical low-field susceptibility $4\pi \chi (T)$ data for three different
samples are shown in Fig. \ref{lowfieldMT}. In this figure, the curves
labeled as $\chi _{{\rm zfc}}$ and $\chi _{{\rm fc}}$ were measured in the
zero-field-cooled and the field-cooled states, respectively. The nominal
superconducting volume fraction, $f_{{\rm nom}}$, in Fig. \ref
{fnomcomparison} was obtained by using the relation $f_{{\rm nom}}=-4\pi
\chi _{{\rm zfc}}(T\ll T_{c})$ and is not corrected for the demagnetization
factors.\cite{Demagnetization} The superconducting volume fractions were
much higher, especially in the high magnetic field region, compared to
previous results, as shown in Fig. \ref{fnomcomparison}. \cite
{Jorgensen93,Er92,Er94,Er97,Tao92,Wiedenhorst98}

The superconducting transition onset in Fig. \ref{lowfieldMT} appears at 43
K, which is the value typically reported for the Sr(La)-112 compound.\cite
{Ikeda93,Jorgensen93,Er91,Er92,Er94,Review,Yao94,Er97} However, we can see
some notable differences from the previous reports. One is a very sharp
transition near $T=43$ K, and another is a well-developed saturation of the
susceptibility at low temperatures, which reflects the formation of a high
quality superconducting Sr(La)-112 phase. To quantify the sharpness of the
magnetization near $T_{c}$, we introduced $T_{{\rm mid}}$ such that $\chi
(T_{{\rm mid}})=0.5\times \chi $($T\sim 0.1\times T_{c}$).\cite{Tmid} The $%
T_{{\rm mid}}$ for sample A is 40.5 K, and $T_{c}-T_{{\rm mid}}$ was only
half the smallest value reported until now, a clear indication of the
sharpness of the superconducting transition.\cite{Jorgensen93} The saturated
values of $4\pi \chi _{{\rm zfc}}$ at low temperatures are about $-1.0$, $%
-1.17$, and $-1.22$ for samples A, B, and C, respectively.

For a superconducting sphere with a radius $R$, $4\pi \chi (T)$ is given by
the Shoenberg formula $-3/2(1-(3/x)\coth x+3/x^{2})$, where $x=R/\lambda _{%
{\rm avg}}(T)$ and $\lambda _{{\rm avg}}(T)$ is the average magnetic
penetration depth, {\em i.e.}, $\lambda _{{\rm avg}}=(\lambda
_{ab}^{2}\lambda _{c})^{1/3}$.\cite{Shoenberg} In the limit of $x\gg 1$, the
absolute value of $-4\pi \chi $ is $\sim 1.5$, which, due to the
demagnetization effect, is 50$\%$ larger than the ideal value.\cite
{Demagnetization} If we take the typical value of $\lambda \sim 2000$ \AA ~
for high-$T_{c}$ cuprates and the grain size $R\simeq 5~\mu $m obtained from
the SEM image, the value of $4\pi \chi $ is estimated to be about $-1.3$,
which is close to the above measured value. Thus, the real superconducting
volume fraction of our sample should be close 100$\%$.

Diamagnetic shielding fraction from $\chi _{{\rm zfc}}$ is basically same
for 100 Oe as it is for 10 Oe, as can be seen in Fig. \ref{lowfieldMT}. This
is quite typical for all of our samples. By measuring the point at which
magnetic hysteresis curve $M$($H$) deviates from linearity, we could
identify the lower critical field $H_{c1}$, and our value was\ above 10$^{2}$
Oe, as are the values for other high-$T_{c}$ cuprates. Previously, it was
claimed that $H_{c1}$ was significantly less than 100 Oe. \cite{Er92}

\begin{figure}[tbp]
\begin{center}
\includegraphics[width=9cm,height=11cm]{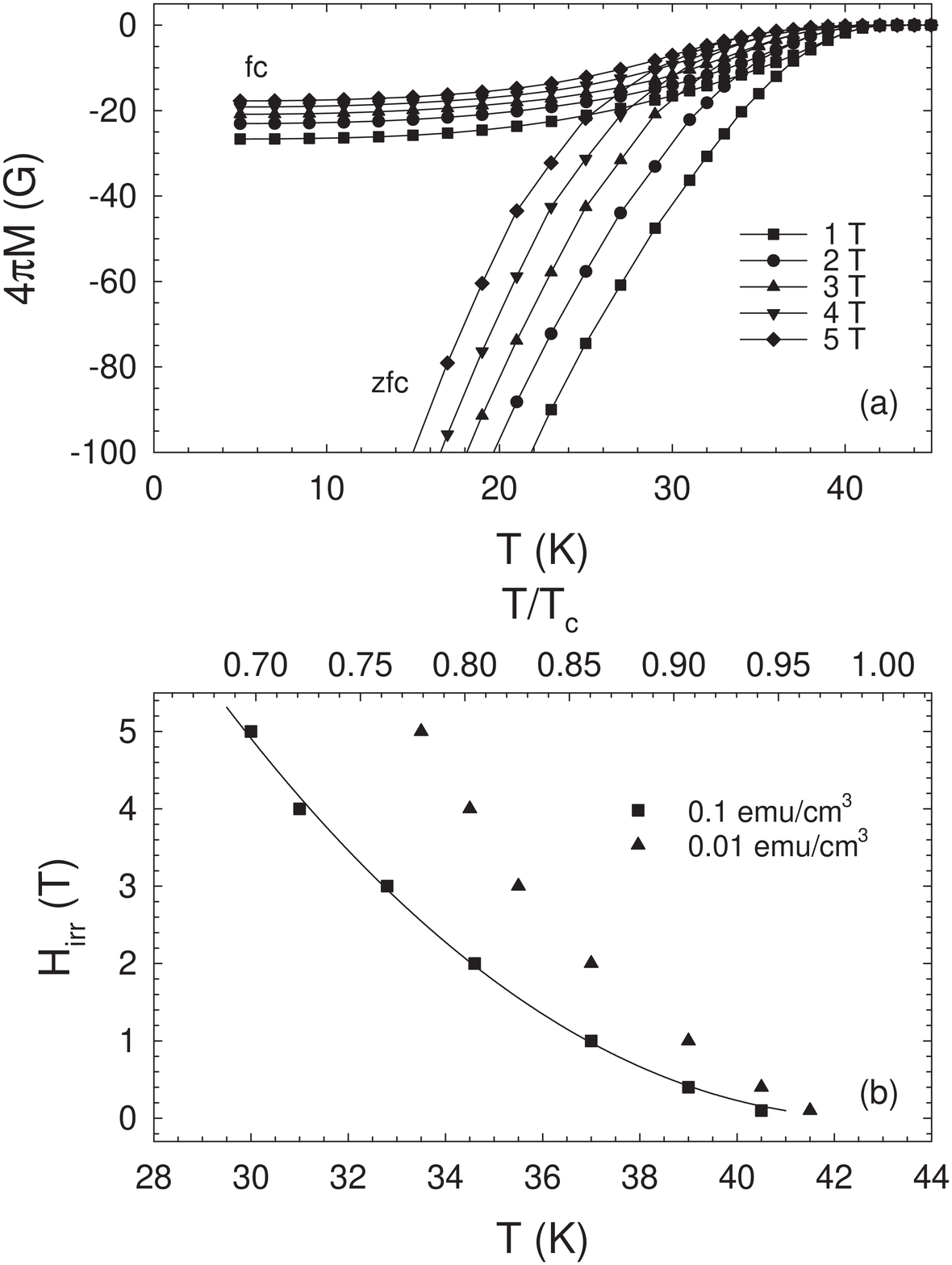}
\end{center}
\caption{$4\protect\pi M(T)$ curves of sample A at fields higher than 1
Tesla and irreversibility field {\it H}$_{irr}$($T$): (a) $4\protect\pi M(T)$
curves at 1, 2, 3, 4, and 5 Tesla, and (b) irreversible field fitted with $%
H_{irr}(T)=H_{o}(1-T/T_{c})^{n}$. The criterion was chosen as $|M_{{\rm zfc}%
}-M_{{\rm fc}}|=0.1$ emu/cm$^{3}$. The error bar in terms of temperature is
less than 0.2 K.\ The fit was excellent with the parameters $H_{o}=55.7$
Tesla, $T_{c}=42.6$ K, and $n=1.99$. The top axis denotes the normalized
temperature $T$/$T_{c}$. The filled triangles were obtained with the
criterion $|M_{{\rm zfc}}-M_{{\rm fc}}|=0.01$ emu/cm$^{3}$.}
\label{Hirr}
\end{figure}

While the low-field magnetization demonstrated a highly enhanced
superconducting volume fraction, the irreversible field $H_{{\rm irr}}(T)$
from the high-field magnetization up to 5 Tesla showed that pinning was very
strong in this compound. In Fig. \ref{Hirr}, the magnetization curves for
fields higher than 1 Tesla and the resulting $H_{{\rm irr}}(T)$ are
presented. The criterion for the reversible point was set as $|M_{{\rm zfc}%
}-M_{{\rm fc}}|=0.1$ emu/cm$^{3}$. The irreversible field was fitted with 
{\it H}$_{irr}$({\it T})= {\it H}$_{0}$(1-{\it T}/{\it T}$_{c}$)$^{n}$. The
best parameters were $H_{0}=55.7$ Tesla, $T_{c}=42.6$ K, and $n=1.99$. The
quite interesting point is that the $H_{{\rm irr}}(T/T_{c})$\ of Sr(La)-112
is about 2 times higher than that of (La,Sr)$_{2}$CuO$_{4}$ even though the
superconducting transition temperature of both compounds are similar.\cite
{Er94} If the criterion is chosen more strictly as $|M_{{\rm zfc}}-M_{{\rm fc%
}}|=0.01$ emu/cm$^{3}$, the irreversible field is increased by a factor of
two.

\begin{table}
\caption{Structural parameters for Sr$_{0.9}$La$_{0.1}$CuO$_{2}$ from
Rietveld refinement using X-ray powder diffraction pattern for sample A. The
values in parentheses are reported ones based on neutron powder diffraction.
Ref 4.}
\label{refinement}
\begin{tabular}{ccc}
{Parameter} &  & Value \\ 
\tableline$a=b$ (\AA ) &  & 3.950 42(3.950 68) \\ 
$c$ (\AA ) &  & 3.410 20(3.409 02) \\ 
$V$ (\AA $^{3}$) &  & 53.219(53.212) \\ 
$\alpha =\beta =\gamma $ &  & 90.000 0 \\ 
Sr,La\tablenotemark[1] & $x=y=z$ & 0.5 \\ 
& $n$ & 1 \\ 
Cu & $x=y=z$ & 0 \\ 
& $n$ & 1 \\ 
O & $x$ & 0.5 \\ 
& $y=z$ & 0 \\ 
& $n$ & 2 \\ 
\tableline Agreement factor &  & Value(\%) \\ 
\tableline R$_{wp}$ (\%) &  & 7.48(16.0) \\ 
R$_{p}$ (\%) &  & 4.68 \\ 
R$_{e}$ (\%) &  & 2.21 \\ 
\tableline Goodness of fit, $S$ &  & 3.3853
\end{tabular}
\tablenotetext[1]{Constraint: n(Sr):n(La)=0.9:0.1.}
\end{table}

\section{summary}

We synthesized the infinite-layer compound Sr$_{0.9}$La$_{0.1}$CuO$_{2}$.
Its improved superconducting properties were confirmed by a structural
analysis and low-field magnetization measurements. The high superconducting
quality was confirmed by the high superconducting volume fractions and the
sharp transition near $T=43$ K. The irreversibility field $H_{{\rm irr}}(T)$%
\ was much higher than that of (La,Sr)$_{2}$CuO$_{4}$, which indicated an
enhanced interlayer coupling between the CuO$_{2}$ planes\ due to a
shortening of the $c$-axis lattice constant. A sufficient quantity of high
quality infinite-layer superconducting samples will ignite research on the
superconductivity in electron-doped infinite-layer superconductors.

\acknowledgments
We are thankful for discussions on the infinite-layer superconductors to K.
Kadowaki, R. S. Liu, and D. Pavuna. We greatly appreciate our valuable
discussions with P. D. Han, D. A. Payne, C. E. Lesher, M. Takano, and A. Iyo
on the general aspects of high-pressure synthesis. This work is supported by
the Ministry of Science and Technology of Korea through the Creative
Research Initiative Program.

\begin{center}
\includegraphics[width=9cm,height=7cm]{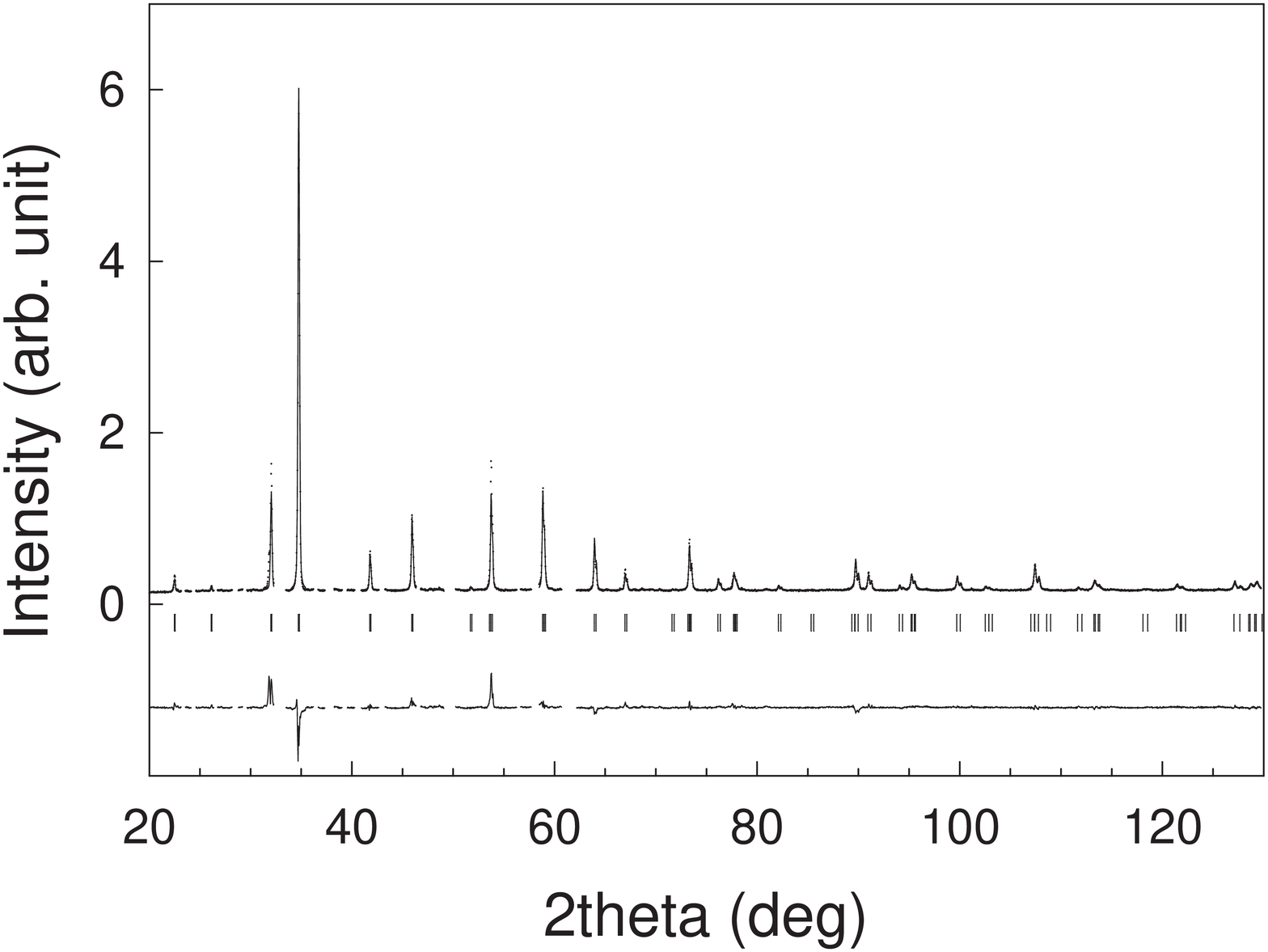}
\end{center}
\begin{figure}[tbp]
\caption{Rietveld refinement of the X-ray powder diffraction pattern of
sample A. The dots are the raw data including background, and the solid line
is the calculated profile. The vertical tick marks below the profile
represent the positions of allowed diffractions in the tetragonal {\it P4/}%
mmm space group. A difference curve (observed pattern minus calculated
pattern) is also plotted at the bottom.}
\label{XRD}
\end{figure}

\vspace*{1cm}      


\begin{references}
\bibitem{Tmid}  This definition of sharpness is more natural for granular
samples grown by using the high-pressure synthesis technique.
\end{references}
\end{document}